\PassOptionsToPackage{dvipsnames, table, x11names}{xcolor}
\documentclass[sigconf]{acmart}

\AtBeginDocument{%
  }

\setcopyright{acmlicensed}
\copyrightyear{2018}
\acmYear{2018}
\acmDOI{XXXXXXX.XXXXXXX}

\acmConference[XXX '25]{XXX}{XXXX--XXXX}{XX, XX}
\acmISBN{978-1-4503-XXXX-X/18/06}

\acmSubmissionID{xxx}

\usepackage{CJK}

\newcommand{\downbad}[1]{\textcolor{Maroon}{\scriptsize \ $\downarrow${#1}}}

\newcommand{\basex}[1]{\textcolor{gray!50}{\scriptsize \ $\uparrow${#1}}}
\usepackage{graphicx}
\usepackage{xcolor}
\usepackage{multirow}
\usepackage{rotating}
\usepackage{multicol}
\usepackage{booktabs}
\usepackage{amsfonts}

\usepackage{amssymb}
\usepackage{pifont}
\usepackage{cleveref}
\usepackage{enumitem}
\Crefname{figure}{Figures}{Figures}
\usepackage{nicefrac}  
\usepackage{xcolor}  
\usepackage{subfigure}
\usepackage{makecell}
\usepackage{xspace}
\usepackage[dvipsnames, table, x11names]{xcolor}
\usepackage{graphicx}
\usepackage{caption}
\usepackage{subcaption}
\usepackage[dvipsnames,table]{xcolor}
\definecolor{myblue}{RGB}{245,245,250}
\newcommand{\name}{\textit{HyperG}\xspace}

\newcommand{\eg}{\emph{e.g., }\xspace}
\newcommand{\ie}{\emph{i.e., }\xspace}

\DeclareMathOperator*{\concat}{\Vert}
\begin{document}

\title{\name: Hypergraph-Enhanced LLMs for Structured Knowledge}


\author{Sirui Huang}
\authornote{Both authors contributed equally to this research.}
\email{sirui.huang@connect.polyu.hk}
\affiliation{%
  \institution{Hong Kong Polytechnic University}
  \city{Hong Kong SAR}
  \country{China}
}
\email{sirui.huang@student.uts.edu.au}
\affiliation{%
  \institution{University Technology Sydney}
  \city{Sydney}
  \country{Australia}
}

\author{Hanqian Li}
\authornotemark[1]
\affiliation{%
  \institution{Hong Kong University of Science and Technology (Guangzhou)}
  \city{Guangzhou}
  \country{China}}

\author{Yanggan Gu}
\affiliation{%
  \institution{Hong Kong University of Science and Technology (Guangzhou)}
  \city{Guangzhou}
  \country{China}
}

\author{Xuming Hu}
 \email{xuminghu@hkust-gz.edu.cn}
\affiliation{%
 \institution{Hong Kong University of Science and Technology (Guangzhou)}
 \city{Guangzhou}
 \country{China}}

\author{Qing Li}
\affiliation{%
  \institution{Hong Kong Polytechnic University}
  \city{Hong Kong SAR}
  \country{China}}

\author{Guandong Xu}
\email{guandong.xu@uts.edu.au}
\affiliation{%
  \institution{University Technology Sydney}
  \city{Sydney}
  \country{Australia}}
  \affiliation{%
  \institution{Education University of Hong Kong}
  \city{Hong Kong SAR}
  \country{China}}


\begin{abstract}
Given that substantial amounts of domain-specific knowledge are stored in structured formats, such as web data organized through HTML, Large Language Models (LLMs) are expected to fully comprehend this structured information to broaden their applications in various real-world downstream tasks. Current approaches for applying LLMs to structured data fall into two main categories: serialization-based and operation-based methods. Both approaches, whether relying on serialization or using SQL-like operations as an intermediary, encounter difficulties in fully capturing structural relationships and effectively handling sparse data. To address these unique characteristics of structured data, we propose \textit{HyperG}, a hypergraph-based generation framework aimed at enhancing LLMs' ability to process structured knowledge. Specifically, \textit{HyperG} first augment sparse data with contextual information, leveraging the generative power of LLMs, and incorporate a prompt-attentive hypergraph learning (PHL) network to encode both the augmented information and the intricate structural relationships within the data. To validate the effectiveness and generalization of \textit{HyperG}, we conduct extensive experiments across two different downstream tasks requiring structured knowledge. Our code is publicly available at: \url{https://anonymous.4open.science/r/HyperG}.
\end{abstract}





\received{20 February 2007}
\received[revised]{12 March 2009}
\received[accepted]{5 June 2009}

\maketitle
\section{Introduction}
%\sr{todo: use ``prompt'' instead of ``instruction''}\sr{todo: citations}
With the advancement of digitalization across various industries, substantial amounts of structured knowledge are stored in tabular formats. This structured knowledge, often containing domain-specific information closely tied to different downstream tasks, complements the general knowledge acquired by Large Language Models (LLMs) during pre-training, thereby enhancing their capability to support downstream queries and reasoning~\cite{cui2024tabular,tan2024struct}.

\begin{figure}[t!]
    \centering
    \includegraphics[width=\columnwidth]{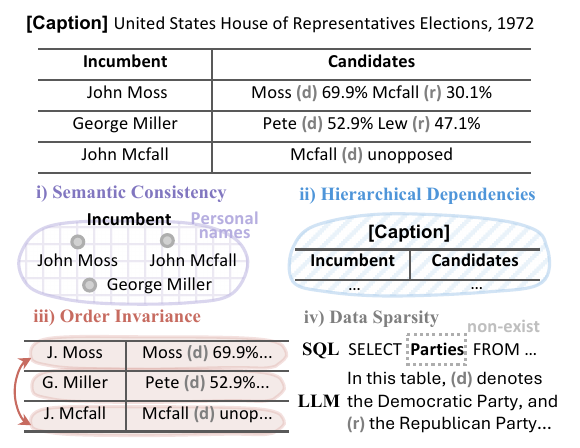}
    % \vspace{-0.31in}
    \caption{An example illustrates the three aspects of the structural relationships in tables: i) Semantic Consistency, ii) Hierarchical Dependencies, and iii) Order Invariance. Additionally, it highlights the data sparsity issue iv), where incomplete data affects SQL queries over the table .}%\sr{the description of toy example in  intro should be modified in accordance with the figure.}}
    \label{fig:toy_example}
% \vspace{-0.20in}
\end{figure}

LLMs, leveraging their sophisticated linguistic capabilities and extensive knowledge base, have been widely utilized as one-/few-shot learners in various structured tasks ~\cite{hegselmann2023tabllm,hanlarge,zhu2023incorporating,kongopentab}. Currently, the approaches for applying LLMs on structured knowledge, including tables, fall into two primary categories: serialization-based~\cite{min2024exploring,hegselmann2023tabllm,jaitly2023towards} and operation-based methods~\cite{ye2023decomposers,jiang-etal-2023-structgpt,wang2024chainoftable,lu2023chameleon}. Serialization-based methods convert structured knowledge into sequences of tokens, enabling the model to process the structured data in conjunction with task descriptions~\cite{min2024exploring,shao2024linearizing}. For example, Hegselmann et al.~\cite{hegselmann2023tabllm} utilize a Table-to-Text model or a LLM as the serializer to convert tables into natural language strings, which are then fed into the LLM along with task descriptions. However, serializing structured data can undermine the inherent structural relationships, especially in larger tables, potentially leading to serve knowledge forgetting and diminished logical coherence during reasoning~\cite{zhang2023ho,li2024snapkv}. Additionally, the serialized formats critically influenced the performance of LLMs~\cite{singha2023tabular}. The operation-based methods extract relevant information from structured data using SQL-like operations based on task requirements, and then incorporate this knowledge into LLMs to generate responses. While the SQL-like operations account for structural relationships, these methods fail to fully harness the extensive knowledge base of LLMs for effective reasoning~\cite{zhu2023incorporating}. As shown in Figure \ref{fig:toy_example} d), the ``(d)'' and ``(r)'' represent the Democratic and Republican parties, respectively. Since the parties are not explicitly listed in a separate column, retrieving information about the political affiliations of incumbents via SQL queries is challenging. However, LLMs can easily interpret this information due to their advanced in-context learning abilities and knowledge base. Therefore, \textbf{structural relationships} and \textbf{data sparsity} are two critical challenges that current methods are not fully account for when reasoning over structured knowledge, which differs fundamentally from the unstructured text inputs LLMs typically handle~\cite{fang2024large}. 

Graphs are structure-aware, making them a natural choice for modeling structural relationships. However, traditional graphs remain insufficient in effectively capturing the group relationships between rows and columns. %Moreover, traditional graph neural networks (GNNs), with their focus on node- or graph-level objectives, struggle to fully understand the task-specific requirements in natural language.
Unlike traditional graphs, where an edge connects only two nodes, a hyperedge in a hypergraph can connect multiple cells nodes in an unordered manner. Hypergraphs consider the \textbf{structural relationships} within tabular data from three aspects: i) Semantic Consistency. Data in the cells of the same row or column in a table generally correspond to a consistent semantic category, allowing LLMs to identify and infer implicit semantic relationships. As illustrated in Figure \ref{fig:toy_example} i), the cells in the ``Incumbent'' column are all personal names. ii) Hierarchical Dependencies. Hyperedges are capable of capturing intricate, higher-order dependencies within structured knowledge, such as the dependencies of the captions, headers, and cells. iii) Order Invariance. Changing word order in natural language can alter meaning, but rearranging rows or columns in a table, \eg{swapping the Moss and McFall rows in Figure \ref{fig:toy_example} iii)}, does not affect the overall semantics. To address the \textbf{sparsity} issue such as the incomplete parties in Figure \ref{fig:toy_example} iv), hypergraphs facilitate high-order information propagation between nodes and hyperedges, thereby supplementing the representations of incomplete cells with information from their neighbors. In addition, the extensive general knowledge embedded in LLMs can be leveraged to address sparse data issues.

%Specifically, we integrate task-specific inquiries for downstream tasks into the information propagation process of the hypergraph neural network, ensuring that the learned knowledge representations are well-aligned with the objectives of the downstream tasks.

%\textcolor{red}{To enhance LLMs' capabilities on structured knowledge, we propose a novel hypergraph-based generation framework. This framework enables LLMs to fully exploit the structural characteristics of tabular data while effectively addressing sparsity issues.}

To enhance LLMs' capabilities on structured knowledge, we propose a novel \textbf{\textit{Hyper}}graph-based \textbf{\textit{G}}eneration framework, namely \textbf{\name}, to facilitate seamless integration of knowledge from structure learning with hypergraph neural networks into LLMs, without losing focus on task-specific requirements. Specifically, \name explicitly guide the LLMs to augment sparse table cells with contextual information. We then construct semantics hypergraphs with the augmented table and introduce a novel Prompt-Attentive Hypergraph Learning (PHL) module that propagates task-specific inquiries in prompts along with embedded semantic knowledge across structures, and train this module jointly with the LLM. Our contributions are concluded as follows:
% \vspace{-0.03in}
\begin{itemize}[leftmargin=*]
    \item \textbf{Towards structural relationship.} We propose \name, which uses hypergraphs to capture the semantic consistency, order invariance, and hierarchical dependencies within structured knowledge, thereby enhancing the LLM’s capability to understand and reason over structured knowledge.
    \item \textbf{Towards data sparsity.} We design a novel hypergraph neural network to tackle the sparsity issue in tabular knowledge by utilizing the generative abilities of LLMs and then facilitates information propagation through hyperedges.
    % \item \textbf{Towards downstream task requirements}
    \item \textbf{Experiments.} We conduct extensive experiments on various downstream tasks involving structured data to validate the effectiveness of our proposed \name framework.
\end{itemize}
\section{Problem Definition}
Aiming to enhance the capability of LLMs in handling knowledge stored in structured data with hypergraphs, in this paper, we consider tables as the structured data sources to illustrate our \name framework. We construct a hypergraph with the structured knowledge in table $\mathcal{T}$. Each table is formally represented as $\mathcal{T}=\{o, h_i, v_{m,n}|0\leq i \leq N,0\leq m \leq M,0\leq n\geq N\}$, where $o$ is the table caption, $h_i$ represents the header for the $i^{th}$ column, $v_{m,n}$ represents the cell at the $m^{th}$ row (denoted as $r_{m}\in\mathcal{R}$), and the $n^{th}$ column (denoted as $c_{n}\in\mathcal{C}$). As depicted in the upper left of Figure \ref{fig:method}, the very upper-left cell is denoted as $v_{0,0}$. The task description prompt $x$, provided in natural language to the LLMs, includes a textual representation of the table $\mathcal{T}$ (\eg{in markdown format}) and the essential inquiry $\omega$ regarding this table, following a specific template, \ie{$\omega\subset x$}. Specifically, the essential inquiry $\omega$ can be claims in fact verification or questions in question answering. For tasks requiring the knowledge stored in $\mathcal{T}$, we aim to help pretrained LLMs (denoted as $LLM(\cdot)$) to understand and extract the structured knowledge relevant to the inquiry $\omega$ stored in $\mathcal{T}$, thereby improving the effectiveness of LLM's final generations. 

% \noindent
% \begin{problem}[Enhancing LLM on Structured Knowledge]
% For each task $t$, given the Large Language Model $LLM(\cdot)$, structured data $\mathcal{T}$, we aim at finding a learner $f$ to extract the relavent structured knowledge from $\mathcal{T}$ to enhance the capability of LLM.
% \end{problem}

\section{Methodology}
\begin{figure*}[htbp]
    \centering
    \includegraphics[width=\textwidth]{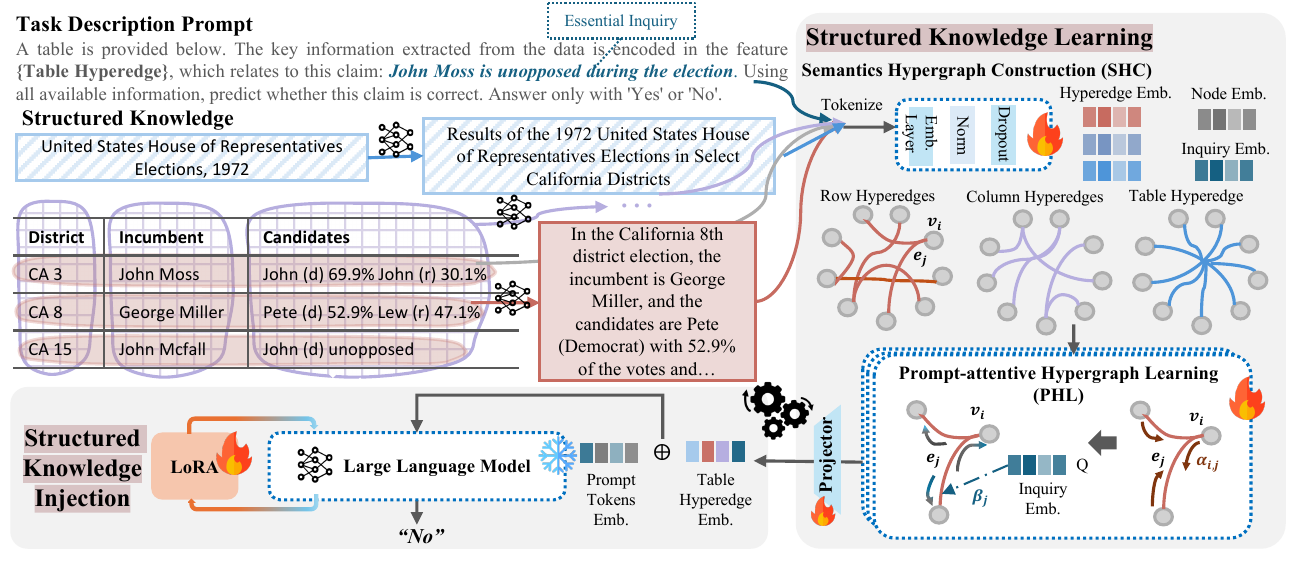}
    % \vspace{-0.2in}
    \caption{An overview of our proprosed \name framework.}
    \label{fig:method}
\end{figure*}

Figure \ref{fig:method} provides an overview of our proposed \name framework, which is designed to enhance the ability of LLMs to handle tasks that require knowledge embedded in structured data.  This section details the workflow of \name, first augmenting the structured data with contextual information, followed by learning and integrating task-relevant structured knowledge into the LLMs to generate answers.

\subsection{Contextual Augmentation}
To address the \textbf{data sparsity} issue caused by missing or incomplete information in cells, \ie{N/A, none, or incompleted descriptions such as the example in Figure \ref{fig:toy_example}}. For each table $\mathcal{T}$, we augment its caption $o$, columns $c\in\mathcal{C}$, and rows $r\in\mathcal{R}$ with contextual information, leveraging the semantics understanding and generative ability of the large language model $LLM(\cdot)$. 

Specifically, as shown in Figure \ref{fig:method}, the caption $o$ ``\textit{United States House of Representatives Elections, 1972}'' is vague, as it does not specify where the elections occurred. After being supplemented with the contextual information from the table, the augmented caption, ``\textit{Results of the 1972 United States House of Representatives Elections in Select California Districts}'', denoted as $\bar{o}$, more clearly illustrates the table's content. As for the sparse cells which contain missing or incomplete data, we utilize the LLM to generate descriptions for each row and column. Formally, for the $m^{th}$ row,

\begin{equation}
        \bar{r}_m = LLM(P_0(o, h_{:}, v_{m,:})
\label{eq:augmented}
\end{equation}
where $\bar{r}_m$ represents the augmented description for the $m^{th}$ row containing cells $v_{m,:}=(v_{m,1}, ..., v_{m,N})$, $h_{:}=(h_1, ..., h_N)$ denotes the $N$ headers of table $\mathcal{T}$, and $P_0(\cdot)$ refers to the template used to prompt the LLM in generating the corresponding augmented summary. Specifically, in this paper, we define the augmentation prompt $P_0$ as: ``\textit{You will be given with the table caption and headers. Please enhance the caption/describe the given row/column corresponding to the table content}.'' The descriptions for the columns $c_n$ in table $\mathcal{T}$ are generated in a similar manner, yeilding $\bar{c}_n$. 

\subsection{Structured Knowledge Learning}
After augmenting the sparse data with contextual information, \name learns \textbf{structural relationships} over knowledge that is structurally stored through two steps: first, by constructing hypergraphs that aligns with the semantics (SHC), and second, by utilizing a novel prompt-attentive neural network for hypergraph learning (PHL). This section will elaborate on how our proposed \name conducts these two steps in detail.

% \vspace{-0.13in}
\subsubsection{Semantics Hypergraph Construction (SHC)} \label{sec:SHC}
This step embeds the semantics of table $\mathcal{T}$ into a hypergraph $\mathcal{G}=\{\mathcal{V},\mathcal{E}\}$, where $\mathcal{V}=\{..., v_{i}, ...\}$ represents the set of node(vertex), and $\mathcal{E}=\{..., e_{j}, ...\}$ represents the set of hyperedges. Each hyperedge connects multiple nodes, $v_i\in \mathcal{N}_{e_j}$ denotes that the node $v_i$ is included in the set of nodes connected by hyperedge $e_j$, while $e_j\in\mathcal{N}_{v_i}$ represents the hyperedge $e_j$ is included in the set of hyperedges which connects node $v_i$. %Here, $i$ and $j$ denote the indices of the nodes and edges across sets $\mathcal{V}$ and $\mathcal{E}$, respectively. 
Each cell $v_{m,n}$ in the table $\mathcal{T}$ is treated as a node, \ie{$v_{i}\in\mathcal{V}$, $|\mathcal{V}|=M\times N$. The rows $r_m\in\mathcal{R}$, columns $c_n\in\mathcal{C}$, and the entire table $\mathcal{T}$ act as hyperedges, leading to three types: row hyperedges $e_\mathcal{R}=\{..., e_{r_m}, ...\}\subseteq\mathcal{E}$, column hyperedges $e_\mathcal{C}=\{..., e_{c_n}, ...\}\subseteq\mathcal{E}$, and table hyperedge $e_\mathcal{T}\subseteq\mathcal{E}$, \ie{$|\mathcal{E}|=M+N+1$}. The connections between the nodes $v\in\mathcal{V}$ and hyperedges $e\in\mathcal{E}$ within the hypergraph $\mathcal{G}$ are represented by an incidence matrix $\mathbf{H}\in\mathbb{R}^{|\mathcal{V}|\times|\mathcal{E}|}$, where each element $h_{i,j}=1$ if node $v_i$ is connected by hyperedge $e_j$, and $h_{i,j}=0$ otherwise. 

%To capture the structured knowledge in table $\mathcal{T}$ for subsequent LLM processing, we learn the semantics of the textual information of cells, row/column descriptions, the caption. 
Specifically, we first tokenize the textual contents of cells, rows, columns, and captions using the BERT~\cite{devlin2019bert} tokenizer. For example, the augmented table caption is transformed into $O$ number of tokens represented by $(\mathbf{t}_{\bar{o},1},\mathbf{t}_{\bar{o},2},..., \mathbf{t}_{\bar{o},O})=Tok_{BERT}(\bar{o})$. The tokens are subsequently passed to an embedding layer, represented as $Emb(\cdot)$, which has an output hidden dimension of $d$ for semantics learning. Layer normalization and dropout layers are implemented in the embedding process to ensure robust generalization capabilities.
\begin{equation}
    \mathbf{h}_{e^{\mathcal{T}}}=Dropout(LN(Emb(\mathbf{t}_{\bar{o},1},\mathbf{t}_{\bar{o},2},..., \mathbf{t}_{\bar{o},O})))
\label{eq:semantics_emb}
\end{equation}
where $\mathbf{h}_{e^{\mathcal{T}}}\in{\mathbb{R}^{d}}$ is the hidden embedding of table hyperedge, $LN(\cdot)$ represents the layer normalization, and $Dropout(\cdot)$ refers to a dropout layer with a dropout rate 0.1. By representing the semantics of each cell content $v\in\mathcal{V}$ as node embedding $\mathbf{h}_{v}$ with Equation (\ref{eq:semantics_emb}), representing the row and column descriptions $\bar{r}_m$ and $\bar{c}_n$ as row/column hyperedges $\mathbf{h}_{e^{\mathcal{R}}}$ and $\mathbf{h}_{e^{\mathcal{C}}}$, and the table hyperedge $\mathbf{h}_{e^{\mathcal{T}}}$, we construct the semantic hypergraph $\mathcal{G}$. Additionally, we also calculate the semantic embedding for the essential inquiry (as highlighted in teal-blue in the task description prompt in Figure \ref{fig:method}), denoted as $\mathbf{h}_{\omega}$ for further learning. 
%To understand the semantics of the structured knowledge in table for further LLM processing, we first learn the semantics embedding of each cells, rows, and columns with a small LM model, for example, $BERT(\cdot)$~\cite{?}, considering previous works suggest that small language models are better at semantics embedding when compared to LLMs ~\cite{}. 

\subsubsection{Prompt-attentive Hypergraph Learning (PHL)}\label{sec:PHL}
Provided with the hypergraph $\mathcal{G}$, we design a prompt-attentive hypergraph neural network to further learn structured knowledge from $\mathcal{G}$. %, ensuring it captures the intricate structural relationships within the data while meeting the task-specific requirements for accurate and context-aware LLM generations. 
In traditional hypergraph learning~\cite{hypergcn,hgnn,hnhn,hcha}, hyperedge embeddings typically do not directly participate in the propagation process; instead, hyperedges primarily serve to connect related nodes, with the focus on node embeddings. In \name, we aim to integrate the semantic embeddings of both nodes and hyperedges during propagation. Since the table cells contain diverse content, while the augmented hyperedge descriptions (\ie{$\bar{o}$, $\bar{r}$, and $\bar{c}$}) are generated by the same LLM and maintain a consistent linguistic style, we apply node-to-edge and edge-to-node propagation using attention scores denoted by $\alpha$ and $\beta$ with distinct designs. Specifically, inspired by ~\cite{allset}, each PHL layer comprises two-step graph attention: first conducts \textit{semantic-aware propagation} from nodes to their connected hyperedges, then \textit{attentively integrate the embedding of the inquiry in the prompt} and propagates from edges to nodes. 

\begin{figure}[tbp]
    \centering
    \includegraphics[width=1.03\columnwidth]{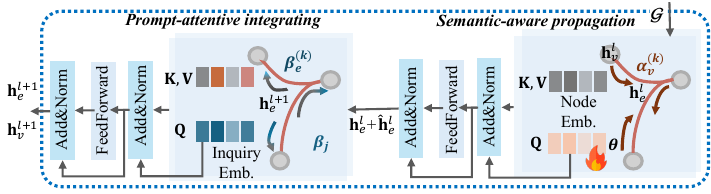}
    % \vspace{-0.1in}
    \caption{The detailed architecture of PHL.}
    \label{fig:phl}
\end{figure}

\textit{Semantic-aware propagating}. Nodes are embedded from the original table cells content with Equation (\ref{eq:semantics_emb}), denoted as $\mathbf{h}_v$. We first propagate the original semantics embedded in $\mathbf{h}_v$ to each connected hyperedge $e$ with the $K$-head hypergraph attention mechanism. The attention score $\alpha_v^{(k)}$ in node-to-edge propagation for node $v$ at the $k^{th}$ head is calculated as follows. 
\begin{equation}
\alpha^{(k)}_v=\text{L-ReLU}(\sum\mathbf{Q}^{(k)}\cdot \mathbf{K}^{(k)}_v)
%=\text{L-ReLU}(\sum_{i=1}^{d}(\mathbf{\theta}_{k,:}\cdot MLP_K(\mathbf{h}_v^{l}))_i)
\label{eq:att_alpha}
\end{equation}
where $\text{L-ReLU}(\cdot)$ denotes the LeakyReLU activation function, the query representation $\mathbf{Q}^{(k)}=\mathbf{W}_{k,:}\in\mathbb{R}^{1\times d}$ is the $k^{th}$ vector of a learnable weight $\mathbf{W}\in\mathbb{R}^{K\times d}$. We use another multi-layer perceptron to learn the key representation of the target node $v$, \ie{$\mathbf{K}^{(k)}_v=MLP_K^{(k)}(\mathbf{h}_v^{l})\in\mathbb{R}^{1\times d}$}. 

Next, the information is propagated from the nodes to their connected hyperedges, formally represented as below.

\begin{equation}
% \begin{aligned}
        \mathbf{h}_{e}^{l,(k)} = \sum_{v\in\mathcal{N}_e}\sigma(\alpha_v^{(k)})\mathbf{V}^{(k)}_v 
\label{eq:mha}
% \end{aligned}
\end{equation}
where $\mathbf{V}^{(k)}_v=MLP_V^{(k)}(\mathbf{h}^{l}_{v})\in\mathbb{R}^{1\times d}$ denotes a multilayer perceptron used to transform the node embedding $\mathbf{h}_v^{l}$ to its value representation, $\mathbf{h}_e^{l}$ and $\mathbf{h}_v^l$ represents the hyperedge $e$ and its connected node $v$, respectively, as input to the $l^{th}$ layer. The $h_v^{l}$ is initialized by the semantics embedding of $v$, \ie{$\mathbf{h}_v^{0}=\mathbf{h}_v\in\mathbb{R}^{1\times d}$}. The softmax function $\sigma(\alpha_{v_i}^{(k)})=\frac{exp(\alpha^{(k)}_{v_i})}{\sum_{v\in\mathcal{N}_e} exp(\alpha^{(k)}_{v})}$ computes the normalized attention score for each node $v_i\in\mathcal{N}_e$.The aggregation $\sum_{v\in\mathcal{N}_e}$ for the embedding of each node $\mathbf{h}_v^{l}$ can be performed using any aggregation function, such as summation.

As shown in Figure \ref{fig:phl}, to increase the representation and generation capability to be compatible with the LLM, the aggregated embedding of hyperedges $\mathbf{h}_e$ are processed using residual connections, normalization, and feed forward layers, following the architecture of  transformer ~\cite{vaswani2017attention}.
\begin{equation}
   \mathbf{\hat{h}}_e^{l} = LN(FF(LN(\concat\nolimits_k\mathbf{h}_e^{l,(k)}+\mathbf{W}))+\concat\nolimits_k\mathbf{h}_e^{l,(k)})
\end{equation}
where $\concat\nolimits_k\mathbf{h}_e^{l,(k)}\in\mathbb{R}^{K\times d}$ represents the concatenation of the outputs of $K$ heads of the multi-head attention mechanism described in Equation (\ref{eq:att_alpha})(\ref{eq:mha}). The $\mathbf{\hat{h}}_e^{l,(k)}\in\mathbb{R}^{1\times d}$ denotes the hyperedge embedding incorporating information propagated from nodes, and is then concatenated to the original semantics embedding of hyperedge $e$, \ie{$\mathbf{h}_e^{l+1} = \mathbf{\hat{h}}_e^{l} + \mathbf{h}_e$, $\mathbf{h}_e^{l+1}\in\mathbb{R}^{1\times 2d}$}. 

\textit{Prompt-attentive integrating}. The original semantics embedding of hyperedge $\mathbf{h}_e$ are embedded from the LLM-augmented descriptions obtained with Equations (\ref{eq:augmented})(\ref{eq:semantics_emb}). To integrate the task requirements described in prompts into hypergraph learning, we adopt the embedding of the essential inquiry $\mathbf{h}_{\omega}$ in the prompt to calculate the attention score $\beta_e^{(k)}$ for edge-to-node propagation, similar to the process in Equation (\ref{eq:att_alpha}) (\ref{eq:semantics_emb}). The essential inquiry can be replaced by any textual information that is of particular relevance or concern to downstream tasks in the prompts.

\begin{equation}
    \beta_v^{(k)} = \text{L-ReLU}(\sum MLP_Q^{(k)}(\mathbf{h}_{\omega}\cdot MLP_K^{(k)}(\mathbf{h}_e^{l+1})))
\label{eq:beta}    
\end{equation}
\begin{equation}
    \mathbf{h}_v^{l,(k)} = \sum_{e\in\mathcal{N}_v} \sigma(\beta_v^{(k)}MLP_V^{(k)}(\mathbf{h}_e^{l+1}))
\label{eq:prompt_emb}
\end{equation}
where $e\in\mathcal{N}_v$ denotes that the set of all the hyperedge $e$ which connects the target node $v$, $MLP_Q^{(k)}(\mathbf{h}_{\omega})\in\mathbb{R}^{1\times d}$ represents the $k^{th}$ vector of the query representation $MLP_Q(\mathbf{h}_{\omega})\in\mathbb{R}^{K\times d}$, computed by a multilayer perceptron based on $\mathbf{h}_{\omega}$. This operation utilizes task requirements to attentively propagate information from hyperedges to nodes. This operation intuitively aligns with the human reasoning process, where relevant rows or columns are identified first, followed by a detailed examination of individual cells when handling tasks that require structured knowledge.

Similarly, the node embedding $\mathbf{h}_v^{l}$ learned from multi-head attentions are further processed with the residual connections, normalization, and feed-forward layers, formally as below.

\begin{equation}
    \mathbf{h}_v^{l+1} = LN(FF(LN(\concat\nolimits_k\mathbf{h}_v^{l,(k)}+MLP_Q^{(k)}(\mathbf{h}_{\omega})))+\concat\nolimits_k\mathbf{h}_v^{l,(k)}))
\end{equation}
where $\concat\nolimits_k\mathbf{h}_v^{l,(k)}$ denotes the concatenation of the outputs of $K$-head attention mechanism described in Equation (\ref{eq:beta})(\ref{eq:prompt_emb}). 

Through the adoptions of semantic-aware propagation and inquiry-attentive integration in each PHL layer, the hypergraph neural network attains a comprehensive understanding of the hierarchical semantics embedded within structured data. This approach ensures semantic consistency, comprehensively captures hierarchical dependencies, and preserves the order invariance property of structural relationships within the knowledge structure.

\subsection{Structured Knowledge Integration}
After completing the hypergraph learning process, the knowledge linked to each cell, along with the columns, rows, and caption, is embedded in the representations of the nodes and hyperedges, which are further integrated into the generation process of LLMs.

\subsubsection{Encoding Structured Knowledge} By connecting each cell node to the table hyperedge formed from the caption, as detailed in Section \ref{sec:SHC}, the hidden embedding $\mathbf{h}_{e_{\mathcal{T}}}$ effectively captures the task-relevant structured knowledge of the entire table $\mathcal{T}$. Therefore, $\mathbf{h}_{e_{\mathcal{T}}}^{L}$ is mapped to the token space of LLM using a projector $\pi$.
\begin{equation}
\mathbf{e}_{\mathcal{T}}=\pi(\mathbf{h}_{e_{\mathcal{T}}}^{L})
\end{equation}
Here, the projected table embedding is denoted as $\mathbf{e}_{\mathcal{T}}\in\mathbb{R}^{d'}$, where $d'$ denotes the dimension of the input tokens for the LLM. In \name, the projector $\pi$ is implemented using two linear layers, with a ReLU activation function in between. The table embedding $\mathbf{e}_{\mathcal{T}}$ is then integrated into the token embeddings $\mathbf{e}_{x,:}=Tok_{LLM}(P_2(x))$ of the task description prompt $x$ at the designated placeholder position labeled ``Table Hyperedge'' in natural language, as highlighted by the bold text in the upper left corner of Figure \ref{fig:method}. 
\begin{equation}
 \mathbf{\hat{e}}_{x}=(\mathbf{e}_{x,:ph-start})\concat(\mathbf{e}_{\mathcal{T}})\concat(\mathbf{e}_{x,ph-end:})
\end{equation}
where $\mathbf{e}_{x,:ph-start}$ denotes all the prompt tokens preceding the placeholder, $\mathbf{e}_{x,ph-end:}$ denotes all the prompt tokens following the placeholder. Additionally,  $\mathbf{\hat{e}}_{x}$ represents the tokens that integrate structured knowledge for further inference in the LLMs.

\subsubsection{Training}
Given the task description, markdown table, and the inquiry $\omega$ in prompt $x$, structured table $\mathcal{T}$, our \name jointly train the prompt-attentive hypergraph learning network with the LoRA~\cite{hu2022lora}. The supervised fine-tuning process can be expressed in terms of the log likelihood loss. Given the input task description prompt $x$ and target output $y$ from the training set $\mathcal{D}$, there is,

\begin{equation}
    \mathbb{E}_{(x,y,\mathcal{T})\in\mathcal{D}}\left[\sum_{t=1}^{T} log p_\theta(y_t|y_{1:t-1}, x, \mathcal{T})\right]
\end{equation}
Here, the conditional probability distribution of the target generation output sentence $y$ given prompt $x$ is represented as $p_\theta(y|x)=\prod_{t=1}^{T}x_\theta(y_t|y_{<t}, x, \mathcal{T})$, where $\theta$ denotes the model parameters and $T$ is the length of the generated sequence.
% \subsection{Training Strategy}
% \subsubsection{Pretraining of PHL Layers}
% \subsubsection{Joint Training of PHL Layers and LoRa}
\section{Experiments}
To validate the effectiveness of \name, we have conduct extensive experiments to answer the following research questions.
\begin{itemize}
    \item \textbf{RQ1}: 
    How does the proposed \name perform compared to state-of-the-art (SoTA) methods when using various LLMs as backbones across different downstream tasks?
    \item \textbf{RQ2}: Is \name scalable to tables of different sizes?
    \item \textbf{RQ3}: How does the proposed \name retain the \textit{Order Invariance} of structural relationships?
    \item \textbf{RQ4}: How does \name retain the \textit{Semantic Consistency} and \textit{Hierarchical Dependencies} of structural relationships?
    \item \textbf{RQ5}: How do the different components of \name contribute to to improving the performance of LLMs in learning from structured knowledge?
    % \item \textbf{RQ3}: How does \name address the complex structural relationships within structured knowledge?
    % \item \textbf{RQ4}: How does the proposed \name integrate structured knowledge with LLMs across different training strategies?

\end{itemize}
\subsection{Experimental Setups}
\subsubsection{Tasks} We validate our proposed \name on two levels of downstream tasks that require fact-checking, and reasoning based on structured knowledge stored in tables.
\begin{itemize}[leftmargin=*]
    % \item \textbf{Table-to-Text (ToT)} To evaluate the structured knowledge understanding ability of LLMs enhanced with our proposed \name, we prompt the LLMs to generate one-sentence descriptions of highlighted sections in tables from the ToTTo~\cite{parikh2020totto} benchmark. The benchmark includes 120,000 training examples, each consisting of a Wikipedia table with a set of highlighted table cells.
    \item \textbf{Table Fact Verification (TFV).} This task aims at assessing the effectiveness of \name in fact-checking over structured knowledge. Specifically, we conduct experiments on the TabFact~\cite{2019TabFactA} benchmark, which contains 16k Wikipedia tables used as evidence for 118k human-annotated claims to explore fact verification with semi-structured knowledge. The ground truth answer for TFV tasks is either ``yes'' or ``no'', signifying whether the given claim is supported or contradicted by the structured knowledge stored in the corresponding table, respectively. 
    \item \textbf{Table Question Answering (TQA).} To validate \name is able to facilitate LLMs to reason over structured knowledge and provide better answers to user input questions, we test on the WiKiTableQuestions~\cite{pasupat2015compositional} dataset, which includes 14,152 examples of open question-answer pairs for training and 4,344 examples for testing. The expected responses for TQA tasks are open-ended answers, which can be in the form of sentences or phrases.
\end{itemize}

% \begin{table}[ht]
%   \caption{The statistics of training sets.}
%   \centering
%   \resizebox{\linewidth}{!}{
%   \begin{tabular}{cc|cccc}
%     \toprule
%     \textbf{Tasks}&\textbf{Answer type}&\textbf{\#Graphs}&\textbf{\#Nodes}&\textbf{\#Hyperedges}\\
%     \midrule
%     &&\textbf{Avg. len}& \textbf{Avg. \#cols}&\textbf{Avg. \#rows}\\
%     \midrule
%     % ToT&one sentence &27491 &86.01 &6.93 &31.82 \\
%     TFV&yes/no &7396 &67.37 &5.77 &13.63 \\
%     TQA&open answer &10375 &65.05 &5.60 &22.34 \\
%     %TODO: dataset statistics (avg #columns, avg #rows, sparsity \\
%     \bottomrule
%   \end{tabular}
% }
% \label{tab:datasets_stats}
% \end{table}

% \begin{table*}[ht]
%   \caption{The statistics of training data.}
%   \centering
%   \scriptsize
%   \resizebox{0.8\textwidth}{!}{
%   \begin{tabular}{ccc|ccc|ccc}
%     \toprule
%     \textbf{Tasks}&\textbf{Datasets}&\textbf{Answer type}&\textbf{\#Graphs}&\textbf{Avg. \#nodes}&\textbf{Avg. \#hyperedges}&\textbf{Inquiry Avg. len}& \textbf{Avg. \#cols}&\textbf{Avg. \#rows}\\
%     \midrule
%     % ToT&one sentence &27491 &86.01 &6.93 &31.82 \\
%     TFV&TabFact&yes/no &7396&78.65&20.39 &67.37 &5.77 &13.63 \\
%     TQA&WikiTableQuestions&open answer &10375 &125.11&28.94 &65.05 &5.60 &22.34 \\
%     \bottomrule
%   \end{tabular}
% }
% \label{tab:datasets_stats}
% \end{table*}
% tabfact&7396 &67.37 &5.77 &13.63 \\
% tqa &10375 &65.05 &5.60 &22.34
% tot &27491 &86.01 &6.93 &31.82

\begin{table}[ht]
  \caption{The statistics of training data.}
  \centering
  \scriptsize
  \resizebox{\columnwidth}{!}{
  \begin{tabular}{cc|cccc}
    \toprule
    \textbf{Tasks}&\textbf{Answer Type}&\textbf{\#Graphs}&\textbf{Avg. \#Nodes}&\textbf{Avg. \#Edges}&\textbf{Inquiry Avg. len}\\
    \midrule
    % ToT&one sentence &27491 &86.01 &6.93 &31.82 \\
    TFV&yes/no &1849&78.65&20.39 &67.37 \\
    TQA&open answer &10141 &125.11&28.94 &65.05 \\
    \bottomrule
  \end{tabular}
}
\vspace{-0.1in}
\label{tab:datasets_stats}
\end{table}
In particular, we follow the preprocessing steps in ~\cite{yin2020tabert} to prepare the training data, and the statistics for our final training data are listed in the table \ref{tab:datasets_stats}. For testing, we retain the original test sets of the two datasets, TabFact~\cite{2019TabFactA} and WikiTableQuestions~\cite{pasupat2015compositional}, to ensure fair comparison with the selected baselines.

\subsubsection{Baselines} We compare our proposed \name against 12 baseline methods, categorized by their different ways of handling tables: operation-based methods~\cite{rajkumar2022evaluating,dater,wang2024chainoftable} that use external operations like SQL queries and serialization-based methods~\cite{zhang2024tablellama,touvron2024llama3,gemma} that transform information in structures into sequences then prompt into the LLMs. In terms parameter sizes, our comparison covers a range of model sizes range from 2 billion to 70 billion parameters. For a fair comparison, we evaluate the operation-based baseline methods~\cite{rajkumar2022evaluating,dater,wang2024chainoftable} using the same backbone LLMs (\ie{\textbf{LLaMA3-8B-Instruct}, \textbf{LLaMA3.2-3B-Instruct}, \textbf{Gemma-2-9B-It}, and \textbf{Gemma-2-2B-It}) as those used for our \name. To reduce the impact of varying instruction-following abilities among different LLMs, we adopt the instruction-tuned verions of all the selected backbone LLMs in our experiments. %First, we select two \textbf{\textit{traditional methods}} fine-tuned on language models.
% \begin{itemize}
%     \item \textbf{TAPAS-Large} ~\cite{herzig2020tapas} encodes both the table structure and the question together to enhance BERT. 
%     \item \textbf{TAPEX-Large} ~\cite{liu2022tapex} uses transformer to generate operations to retrieve answers from tables.
% \end{itemize}
% For both the TAPAS~\cite{herzig2020tapas} and the TAPEX~\cite{liu2022tapex}, we use the versions of them fine-tuned on TabFact~\cite{2019TabFactA} and WikiTableQuestions~\cite{pasupat2015compositional} in the two tasks TFV and TQA, respectively. 
%Specifically, we select four LLM accross different parameter scales, ranging from 2 billion to 27 billion. In particular, we apply our \name to three of these LLMs to evaluate its generalization capability.

\begin{table*}[htbp]
\scriptsize
\centering
\caption{Comparison of the performance of our \name\ and 13 baseline methods based on varying parameter sizes, where the TFV and TQA tasks are evaluated with respect to Acc. and Denot. Acc., respectively. The first group of methods prompts LLMs with serialized tables, while the methods in each of the last four groups use the same backbone LLMs. The best and second-best results are marked with bold and underline, respectively.}
\resizebox{\textwidth}{!}{
\begin{tabular}{llccccccc}
\toprule
 \multirow{2}{*}[-0.5ex]{\textbf{Methods}} &\multirow{2}{*}[-0.5ex]{\textbf{Backbones}} &\multicolumn{3}{c}{\textbf{\;\;TFV}} & \multicolumn{4}{c}{\textbf{\;\;TQA}} \\
\cmidrule(l){3-5} \cmidrule(l){6-9} 
 &&\bf Acc.   &\bf Prec.   &\bf F1   &\bf Denot. Acc. &\bf ROUGE-1   &\bf ROUGE-2   &\bf ROUGE-L     \\
 \midrule
 \midrule
 Gemma-2-2B-It~\cite{gemma} &- &59.80   &60.55   &58.54   &31.88   &39.68   &17.81   &39.60    \\
LLaMA3.2-3B-Instruct~\cite{touvron2024llama3} &- & 54.90   &56.66   &54.89   &24.77   &35.11   &16.62   &35.07    \\
TableLlama~\cite{zhang2024tablellama} &LLaMA2-7B &70.04 &71.27 &69.39 &24.63 &28.07 &13.95 &27.98 \\
LLaMA3-8B-Instruct~\cite{touvron2024llama3} &- &66.29   &66.29   &66.28   &37.85   &47.58   &21.42   &47.49   \\
Gemma-2-9B-It~\cite{gemma} &- &75.00   &75.20   &74.99  &46.85  &55.73   &24.99   &55.73    \\
Gemma-2-27B-It~\cite{gemma} &- &\underline{76.50} &\underline{76.29} &\underline{75.94} &\underline{53.96} &\underline{61.39} &\underline{27.88} &\underline{61.31} \\
LLaMA-3.1-70B-Instruct~\cite{touvron2024llama3} &-&\textbf{79.16}  &\textbf{79.67} &\textbf{79.12}  &\textbf{55.71}  &\textbf{64.71} &\textbf{29.14}&\textbf{64.70}\\
GPT-4o-mini~\cite{gpt4ominiurl} &- &71.09 &75.58 &70.05  &21.22 &36.42  &19.73 &36.43 \\
GPT-3.5-Turbo~\cite{openai2023gpt35turbo} &- &62.03  &70.86 &58.27 &19.96 &33.69 &18.67 &33.63  \\
\midrule
\midrule
  Text-to-SQL~\cite{rajkumar2022evaluating} &LLaMA3.2-3B-Instruct & 57.80  &58.33 &56.33 &28.84 &35.22 &12.89 &34.79\\
  Dater~\cite{dater} &LLaMA3.2-3B-Instruct & 60.03 &58.39 &59.10 &33.93 &39.18 &13.58 &39.05\\
  CHAIN-OF-TABLE~\cite{wang2024chainoftable} &LLaMA3.2-3B-Instruct & \underline{61.09}  &\underline{60.49} &\underline{60.49} &17.14 &26.81 &12.97 &26.67 \\
LoRA~\cite{hu2022lora} &LLaMA3.2-3B-Instruct &55.21 &57.84  &57.34 &\underline{36.33} &\underline{42.51} &\underline{19.71} &\underline{42.51}\\
  % + Baseline1(hytrel) &51.91 \up{0.0} &52.4 \up{0.0}  &51.91 \up{0.0} &- \up{0.0} &- \up{0.0} &- \up{0.0} \\
 \rowcolor{myblue} \textbf{\name (Ours)} &LLaMA3.2-3B-Instruct &\textbf{61.95}  &\textbf{61.95}  &\textbf{61.93} &\textbf{48.50} &\textbf{54.50} &\textbf{25.80} &\textbf{54.47} \\ 
 \midrule
Text-to-SQL~\cite{rajkumar2022evaluating} &LLaMA3-8B-Instruct & 69.72  &67.20 &69.63 &39.24 &48.28 &20.07 &47.79 \\
  Dater~\cite{dater} &LLaMA3-8B-Instruct & 73.37  &72.42 &73.59 &48.30 &51.74 &18.37 &51.54 \\
  CHAIN-OF-TABLE~\cite{wang2024chainoftable} &LLaMA3-8B-Instruct &\underline{78.06}  &\underline{78.08} &\underline{78.06} &36.97 &46.09 &19.39 &46.1  \\
 LoRA~\cite{hu2022lora} &LLaMA3-8B-Instruct & 66.32  &67.16  &63.48 &\underline{49.65} &\underline{56.78} &\underline{25.76} &\underline{56.76} \\
  \rowcolor{myblue} \textbf{\name(Ours)} &LLaMA3-8B-Instruct &\textbf{79.14}   &\textbf{80.59}   &\textbf{78.95}  & \textbf{55.39}  & \textbf{61.45} &\textbf{27.61} &\textbf{61.37} \\
\midrule 
\midrule
 Text-to-SQL~\cite{rajkumar2022evaluating} &Gemma-2-2B-It & 51.28 &51.15 &52.81 &34.62 &45.89 &18.39 &44.71\\
Dater~\cite{dater} &Gemma-2-2B-It & 55.68  &54.82 &57.55 &\underline{40.95} &\textbf{55.30} &\underline{19.13} &\textbf{55.11} \\
CHAIN-OF-TABLE~\cite{wang2024chainoftable} &Gemma-2-2B-It & 57.66  &\underline{61.49} &57.56 &38.23 &45.94 &18.52 &45.81 \\
LoRA~\cite{hu2022lora} &Gemma-2-2B-It &\underline{59.24} & 59.81  &\underline{58.02} &25.15  &33.04 &15.59 &32.97\\
 \rowcolor{myblue} \textbf{\name (Ours)} &Gemma-2-2B-It &\textbf{60.64}  &\textbf{61.78} &\textbf{60.64} &\textbf{41.80} &\underline{47.70} &\textbf{22.01} &\underline{47.70}\\ 
 \midrule
 Text-to-SQL~\cite{rajkumar2022evaluating} &Gemma-2-9B-It & 70.18  &71.21 &72.03 &50.88 & 54.46 &19.71 &51.53 \\
  Dater~\cite{dater} &Gemma-2-9B-It & 72.88  &71.60 &73.31 &\underline{57.94} &\underline{61.95} &22.91 &61.64\\
  CHAIN-OF-TABLE~\cite{wang2024chainoftable} &Gemma-2-9B-It & 61.55 &\textbf{79.59} &71.77 &50.83 &61.92 &\textbf{28.16} &\underline{61.74} \\
LoRA~\cite{hu2022lora} &Gemma-2-9B-It & \underline{75.56}  & 75.69 &\underline{75.56} &31.72 &53.42  &24.30 &53.39 \\
 \rowcolor{myblue} \textbf{\name (Ours)} &Gemma-2-9B-It &\textbf{79.14} &\underline{79.20}  &\textbf{79.13}  &\textbf{58.54} &\textbf{62.60} &\underline{28.07} &\textbf{62.56} \\ 
\bottomrule
\end{tabular}
}
\label{tab:main}
\vspace{-0.5em}
\end{table*}

\begin{itemize}[leftmargin=*]
    \item \textbf{GPT-3.5-Turbo}~\cite{openai2023gpt35turbo} is an advanced language model from the GPT-3 family by OpenAI, distinguished by its exceptional balance between cost-efficiency and performance, offering faster inference and lower deployment costs.
    \item \textbf{GPT-4o-mini}~\cite{gpt4ominiurl} is a lightweight variant of GPT-4 by OpenAI, offering strong performance with lower computational demands, ideal for resource-constrained applications.
    \item \textbf{LLaMA3.1-70B-Instruct} ~\cite{touvron2024llama3} is a super large model with 70 billion parameters, offering improved performance on more complicated and long-context reasoning.
    \item \textbf{LLaMA3-8B-Instruct} ~\cite{touvron2024llama3} is an instruction-tuned model of the LLaMA3 series with 8 billion parameters, optimized for better instruction understanding and generation. 
    \item \textbf{LLaMA3.2-3B-Instruct} ~\cite{touvron2024llama3} is a relatively small model in the LLaMA3.2 series, fine-tuned with 3 billion parameters, and specifically designed for those tasks requiring rapid responses under limited computational resources.
    \item \textbf{Gemma-2-It} ~\cite{gemma} is fine-tuned on Gemma-2 with user interactions, focusing on task-specific adaptability while ensuring efficiency through knowledge distillation from the very large model. In this paper, we use the 2B, 9B, 27B variants of Gemma-2-It.
    \item \textbf{TableLlama} ~\cite{zhang2024tablellama} adopts LongLoRA to finetune on a dataset that includes a diverse range of serialized tables and the corresponding natural language task instructions.
    \item \textbf{Text-to-SQL}~\cite{rajkumar2022evaluating} designs in-context samples to instruct LLMs in generating SQL queries for answering questions. 
    \item \textbf{Dater} ~\cite{dater} leverages LLMs to decompose the task into multiple sub-tasks, utilizing SQL queries to address each sub-task.
    \item \textbf{CHAIN-OF-TABLE} ~\cite{wang2024chainoftable} prompts LLMs through in-context learning to iteratively produce operations and update the table, thereby constructing a reasoning chain in a structured format.
    \item \textbf{LoRA}~\cite{hu2022lora} is a widely used technique for efficiently fine-tuning LLMs by updating a small number of low-rank weights.  
\end{itemize}

\subsubsection{Evaluation Protocol} We evaluate the generation of LLMs enhanced by our proposed \name framework with respect to the different tasks. For the TFV task, where the answers are either ``yes'' or ``no'', we employ \textbf{accuracy}, \textbf{precision}, \textbf{recall}, and \textbf{F1 score} as the evaluation metrics. To mitigate the impact of option bias~\cite{pezeshkpour-hruschka-2024-large,zheng2024large} in LLMs, we use a weighted version of all these metrics. For the TQA task, where responses may take the form of sentences or phrases, we adopt the following natural language evaluating metrics.
\begin{itemize}[leftmargin=*]
    \item \textbf{Denotation Accuracy (Denot. Acc.)}~\cite{deacc}, following ~\cite{jiang-etal-2023-structgpt,wang2024chainoftable}, measures how closely a response matches the ground truth answer, regardless of the order of phrases in the answers.
    % \item \textbf{Exact Match (EM)} measures the percentage of correct responses that exactly match the ground truth answers.
    \item \textbf{ROUGE-N} measures the similarity between the LLM-generated responses and the ground truth answers by comparing overlapping n-grams, used to evaluate text summaries or translations by quantifying shared word sequences. In this paper, we report both ROUGE-1 and ROUGE-2 scores.
    
    \item \textbf{ROUGE-L} evaluates the similarity between the LLM-generated responses and the ground truth answers by identifying the longest common subsequence (LCS) and is used to assess the fluency and coherence of the generated text.
    
    % \item \textbf{BLEU} evaluates the quality of LLM-generated responses by comparing the n-grams in the output to those in the ground truth answers. It computes precision for n-grams and includes a brevity penalty to discourage excessively short responses, making it a widely used metric in machine translation tasks.

\end{itemize}

\subsubsection{Implementations Details} For \name, we explore the learning rates for the LoRA module within the range of \{5e-5, 1e-5, 5e-6\}, while applying scaling factors for the learning rate in the PHL module (\ie{the novel hypergraph neural networks}) and the projector from \{1, 10, 20\}. We search batch sizes from \{8, 16, 32\}, and conduct experiments over 1 to 4 epochs, utilizing an early stopping strategy. Specifically, for the LoRA module, we fine-tune the Query, Key, and Value projectors with a rank of 8, a LoRA alpha of 32, and a dropout rate of 0.1. For the selected baseline models, we adopt the optimal configurations from the HuggingFace\footnote{https://huggingface.co/models} and accelerate inference with vllm 0.5.4\footnote{https://github.com/vllm-project/vllm}. All experiments in this paper are conducted on two NVIDIA A800-SXM4-80GB GPUs. For further details, please refer to our publicly released code linked in the Abstract Section.

\subsection{Task Performance (RQ1)}
Table \ref{tab:main} presents a comparison of the performance of our \name with 13 baseline methods, encompassing both serialization-based methods~\cite{openai2023gpt35turbo,gpt4ominiurl,zhang2024tablellama,gemma,touvron2024llama3,hu2022lora} and operation-based methods~\cite{rajkumar2022evaluating,dater,wang2024chainoftable}. We evaluate their capabilities in structured knowledge using the TFV task on the TabFact dataset~\cite{2019TabFactA} and the TQA task on the WiKiTableQuestion dataset~\cite{pasupat2015compositional}. In Table \ref{tab:main}, the first group consists of serialized-based methods utilizing various LLMs, while the last four groups compare the performance of our \name with both state-of-the-art operation-based and serialized-based methods across four backbone LLMs. The following observations can be drawn from the performance results in Table \ref{tab:main}.
%Table \ref{tab:main} presents the performance of our \name and various LLMs, encompassing both serialization-based ~\cite{openai2023gpt35turbo,gpt4ominiurl,zhang2024tablellama,gemma,touvron2024llama3} and operation-based~\cite{rajkumar2022evaluating,dater,wang2024chainoftable} models, which are tested on the TFV task with the Tabfact~\cite{2019TabFactA} dataset and the TQA task utilizing the WiKiTableQuestion~\cite{pasupat2015compositional} dataset. Moreover, we apply our \name to four commonly used LLMs, with parameter sizes ranging from 2 billion to 9 billion, to evaluate its generalization capabilities. The following observations can be drawn from the performance results in Table \ref{tab:main}. 

\begin{itemize}[leftmargin=*]
\item \textbf{Our \name outperforms both the operation-based and serialization-based methods based on LLMs.} It can be found in Table \ref{tab:main} that our \name consistently achieve competing performances across both the TFV and TQA tasks. In general, operations-based methods~\cite{rajkumar2022evaluating,dater,wang2024chainoftable} achieves better outperform the methods~\cite{openai2023gpt35turbo,gpt4ominiurl,zhang2024tablellama,gemma,touvron2024llama3,hu2022lora} of merely prompting LLMs with serialized information, even when the models have been fine-tuned on structural data~\cite{zhang2024tablellama}. This highlights the importance of maintaining structures when reasoning about questions related to structured data. Our proposed \name utilizes hypergraphs to encode structural information, complementing the powerful natural language capabilities of LLMs. It demonstrates an average improvements of 1.73\% and 2.43\% in accuracy on TFV and TQA, respectively, when compared to the second-best performances in each group. Upon reviewing the response examples, we found that CHAIN-OF-TABLE~\cite{wang2024chainoftable} encounters difficulties in TQA due to the loss of the question while reasoning over extended chains.
    \item \textbf{Our \name narrows the performance gap between large and small LLMs, requiring only a modest number of additional parameters.} Table \ref{tab:main} shows that instruction-tuned LLMs~\cite{gemma,touvron2024llama3,zhang2024tablellama} with larger parameter sizes achieve better performance with serialization compared to their smaller counterparts. For instance, LLaMA-3.1-70B-Instruct achieves an accuracy of 79.16\%, whereas LLaMA-3-8B-Instruct attains only 66.29\%. Nevertheless, our \name intergrated with LLaMA-3-8B-Instruct, which adds approximately \textbf{189M} parameters (roughly one-tenth of the parameter difference between LLaMA-3-8B-Instruct and LLaMA-3.1-70B-Instruct),  achieves performance comparable to the larger model across both tasks. \name provides average improved accuracy of 6.22\% and 15.72\% on the four backbone models regarding the two tasks, respectively. Similarly, \name based on the Gemma-2-9B-It surpasses its 27B variant by 2.64\% and 2.58\% in the TFV and TQA tasks with respect to accuracy, respectively. This superiority is generalizable from the TQA task to the TFV task, and is attributed to \name's ability in encoding enriched structured knowledge, enabling LLMs to produce more accurate answers. Additionally, the results in Table \ref{tab:main} further demonstrate that our \name enhances LLMs across parameters sizes ranging from 2B to 9B.

\item \textbf{Our \name delivers improvements in performance and training efficiency compared to other SFT methods.} As Supervised Fine-Tuning (SFT) is required in our \name, we compare it to the other two SFT baseline methods: LoRA~\cite{hu2022lora} and TableLlama~\cite{zhang2024tablellama}. First, \name demonstrates significant improvements, achieving an average enhancement of 6.13\% in accuracy and 15.22\% in denotation accuracy~\cite{deacc} over LoRA for the TFV and TQA tasks, respectively.  While LoRA is widely recognized as an efficient tool for instruction tuning, it proves less effective when applied to smaller-scale LLMs, such as Gemma-2-2B-It, particularly for reasoning over serialized structured data. This limitation highlights the challenges of adapting LoRA to tasks requiring nuanced structural understanding. Furthermore, when compared to TableLlama~\cite{zhang2024tablellama}, which is fine-tuned on a benchmark involving serialized structured knowledge, \name provides a more efficient solution by fine-tuning on 189M additional parameters with very limited training data (see Table \ref{tab:datasets_stats}). These observations reinforce our earlier assertion that serialization-based methods can undermine the preservation of structures, further highlighting the necessity of \name in enhancing LLMs to fully utilize such knowledge for improved reasoning.
   
    % \item \textcolor{red}{\textbf{Our \name delivers more substantial enhancements to larger LLMs with more parameters than to smaller ones.}} The last four parts in Table \ref{tab:main} demonstrates the performance of our proposed \name applied on various LLMs, with parameters ranging from 2B to 9B. It can be found that the information of structured knowledge encoded by the PHL module (\ie{the proposed hypergraph neural network}) in \name provides more enhancements on the larger LLMs compared to their small-scale variants. For example, \name achieves a 10.91\% improvement in accuracy using the LLaMA3-8B-Instruct as the base model, compared to a 7.05\% improvement in accuracy on the Gemma-2-9B-It in the TFV task. We attribute the observed attenuation to the small-scale model being distilled from the larger one, which results in more rigid parameters and complicates the integration of structured knowledge.

\end{itemize}

\subsection{Order Invariance (RQ2)}
\begin{figure}[t!]
    \centering
  \begin{minipage}[t]{0.49\linewidth} %
    \subfigure[Shuffle Rows]{
        % \label{fig:barchart1}
        \includegraphics[width=\linewidth]{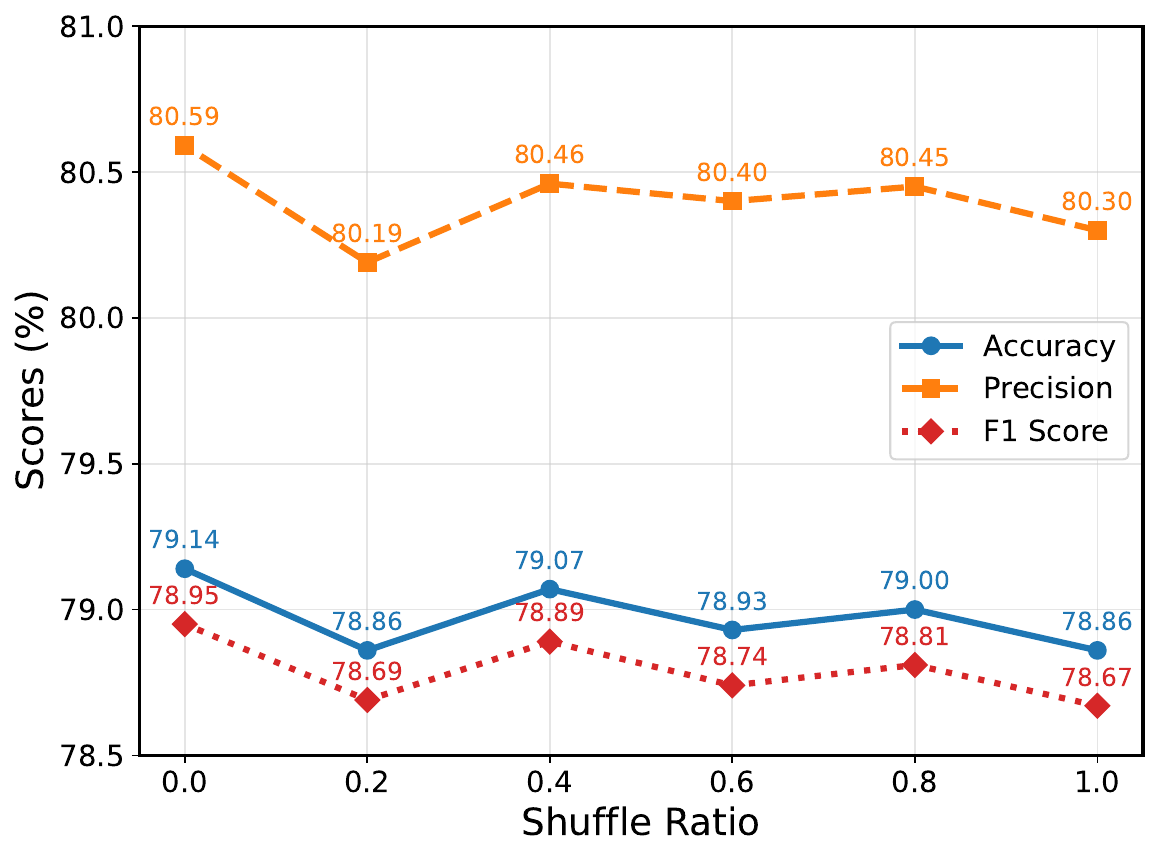}
    }  
  \end{minipage}
  \begin{minipage}[t]{0.49\linewidth} %
    \subfigure[Shuffle Columns]{
        % \label{fig:barchart2}
        \includegraphics[width=\linewidth]{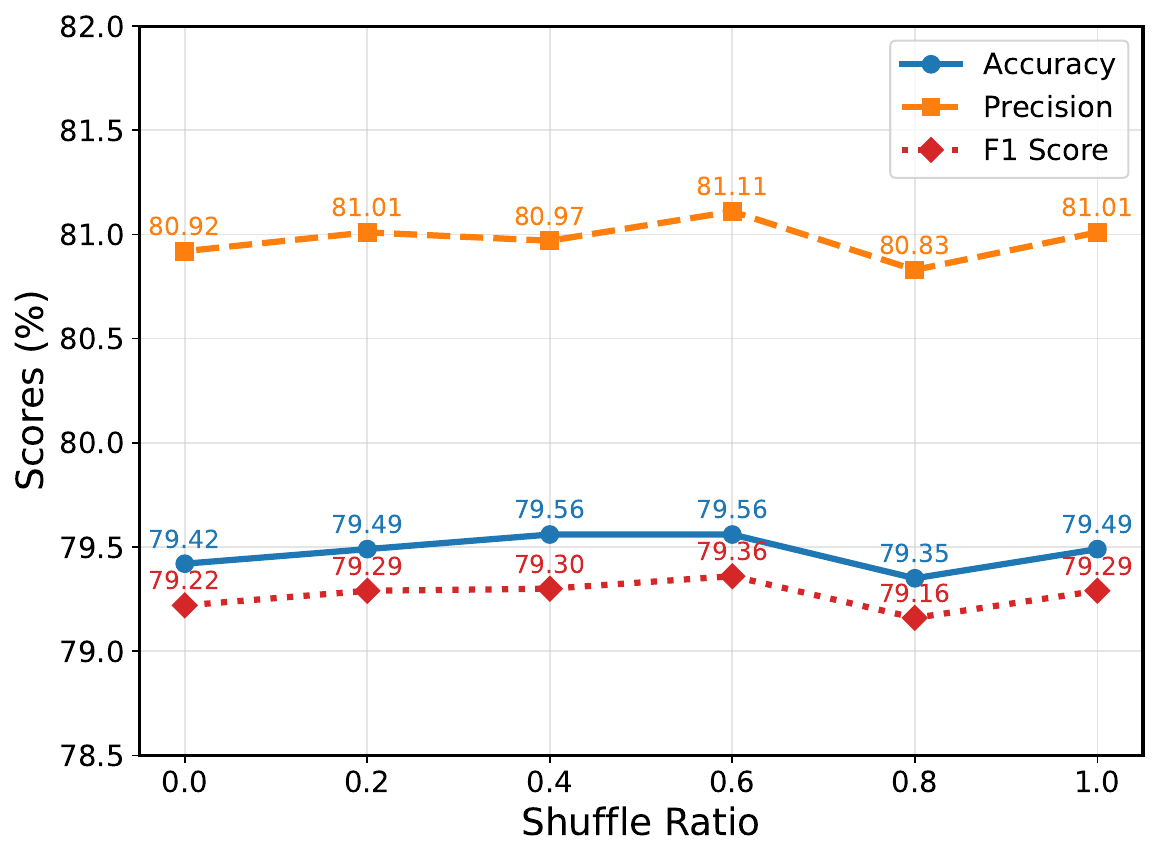}
    }  
  \end{minipage}
\vspace{-0.1in}
\caption{Performances of \name under different variances of order simulated by shuffling.}
\vspace{-0.2in}
\label{fig:order}
\end{figure}
In contrast to natural language, where changes in word order can modify the meaning of a sentence, rearranging rows or columns in a table does not affect its meaning. In \name, this invariance is handled with hyperedges, which represent rows and columns, are inherently unordered within the structure of hypergraphs. To assess how our proposed \name framework helps the LLM maintain the \textit{Order Invariance} of structural relationships, we shuffle the rows in the test data to evaluate \name's robustness to order variations. 

Specifically, we randomly sampled a subset of tables from the TFV testing set and performed shuffling of the rows and columns respectively within each sampled table to introduce variability and evaluate the performance of our proposed \name. Figure \ref{fig:order} displays the performances of \name which uses LLaMA3-8B-Instruct as the backbone model, across different shuffle ratios. The x-axis represents the sampling ratio, while the y-axis indicates performance scores with respect to accuracy, precision, and F1 score. As decipted in Figure \ref{fig:order}, \name framework demonstrates stable performance despite variations in row and column order. The accuracy variance of 0.0109 for row shuffling and 0.00043 for columns shuffling, respectively. This stability underscores the robustness of \name in maintaining structural representation integrity from the perspective of order invariance, thereby validating our previously stated rationale for employing hypergraphs.

\subsection{Semantic Consistency and Hierarchical Dependencies (RQ3)}
\begin{figure}[t!]
    \centering
  \begin{minipage}[t]{\linewidth} %
    \subfigure[\texttt{Case 1}. Information about the baseball teams at Bosse Field.]{
        \includegraphics[width=\linewidth]{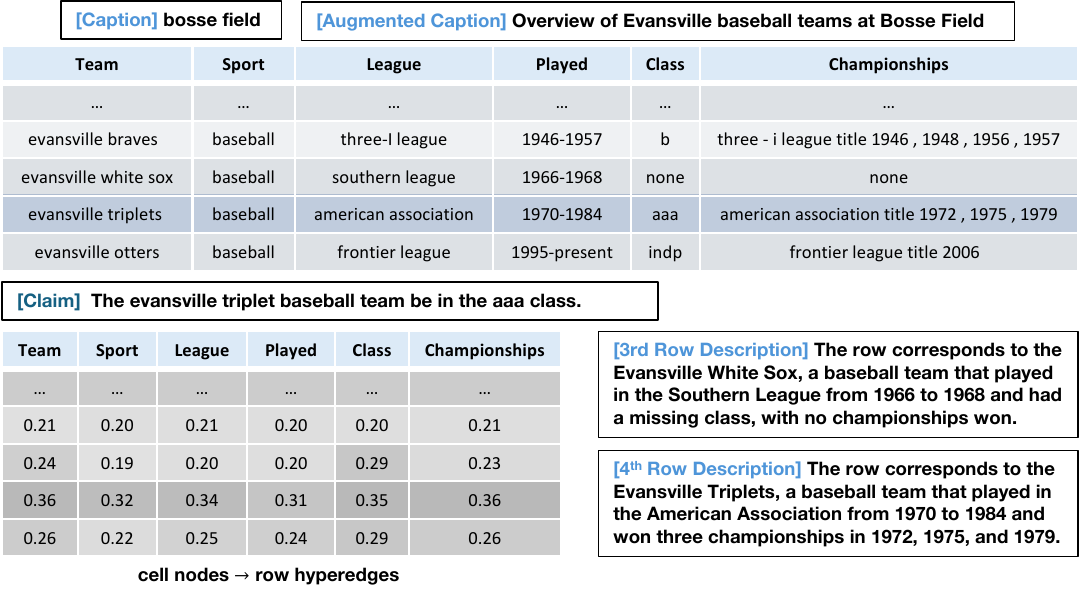}
    }  
  \end{minipage}\\
  \begin{minipage}[t]{\linewidth} %
    \subfigure[\texttt{Case 2.} The results of the 1961 Victorian Football League (VFL).]{
        \includegraphics[width=\linewidth]{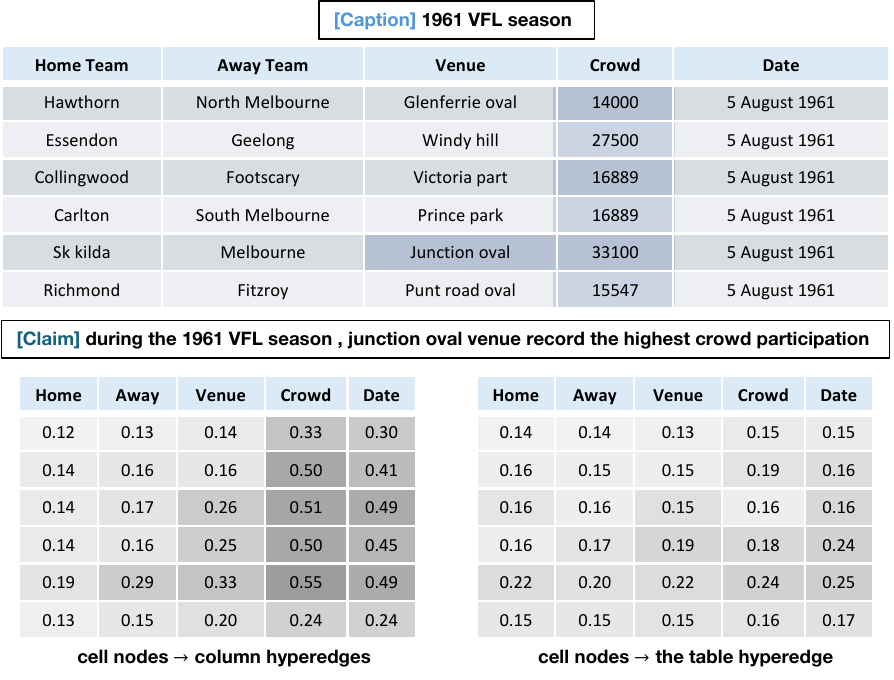}
    }  
  \end{minipage}
\vspace{-0.1in}
\caption{Visualization of the weights between cell nodes and different hyperedges in two random cases.}
\label{fig:case_study}
\end{figure}
Beyond quantitative metrics, we also conduct qualitative evaluations to investigate whether \name retains semantic consistency and hierarchical dependencies of structural relationships during reasoning. In Figure \ref{fig:case_study}, we randomly selected two cases and visualized the attention weights between each cell node and the hyperedges associated with the claim's content. This visualization provides insights into how \name prioritizes and propagates information between table elements and their relevance to the given queries/claims, specifically demonstrating its ability to maintain semantic consistency and hierarchical dependencies.

The semantic consistency in structural realtionships suggests that cells in the same column are similar in semantics. In \texttt{Case 1}, the claim pretains to the class of a team named ``evansville triplets''. \name first augmented each row with easy-to-understand natural language descriptions based on the row context, as shown in the bottom right of Figure \ref{fig:case_study}. This augmentation enables \name to better interpret cells with specific missing values (\eg{``none'' in the ``Class'' columns}), leading to similar weights for these cells and their counterparts within the same column. In \texttt{Case 2}, the claim queries about the highest crowd participation, which requires examining the ``Crowd'' column to identify the largest number. Though the crowd numbers for these teams vary, the weights assigned to the column hyperedges are similar thanks to the augmented column descriptions (omitted here) and the Semantics Hypergraph Construction (SHC) in \name. This is more evident in the weight matrix associated with the table hyperedge shown in the right bottom of \ref{fig:case_study} (b). Even though the venue names are quite different, the cells in the "Venue" column share similar weights.

The Hierarchical Dependencies refers to the hierarchy across cells, columns, rows, and the whole table. As depicted in Figure \ref{fig:case_study}, the attention of \name is primarily focused on the cells and the rows/columns related to the claim. This focus extends from cell nodes to row/colum hyperedges, and then the table hyperedges, gradually diminishing in intensity. For example, in Figure \ref{fig:case_study} (b), the weights assigned to the queried cell ``Junction oval'', the evidence cell ``33100'', and the relevant column ``Crowd'' exceed the average weights of 0.27 and 0.16 in the weight matrices corresponding to the column hyperedges and the table hyperedge, respectively. This demonstrates how the attention mechanism spans the hierarchical structure, emphasizing specific elements within the table.

\subsection{Ablation Study (RQ4)}
\begin{table}[t!]
\centering
\scriptsize
\caption{The ablation study results of \name using LLaMA3-8B-Instruct as the base model on the TFV task. \textcolor{Maroon}{Red} signifies degradation in percentage.}
\resizebox{\columnwidth}{!}{
\begin{tabular}{lllll}
\toprule
 \multirow{2}{*}[-0.5ex]{\textbf{Methods}} & \multicolumn{4}{c}{\textbf{\;\;TFV}} \\
\cmidrule(l){2-5}
 &\bf\%Acc. &\bf\%F1 &\bf\%Prec. &\bf\%Recall \\
\midrule
 \rowcolor{myblue} \textbf{\name (Ours)} &77.20  \basex{0.00} &77.46 \basex{0.00} &79.98 \basex{0.00} &77.20  \basex{0.00}  \\ 
  % w/o LoRA &- \downbad{0.0} &- \downbad{0.0} &- \downbad{0.0} &- \downbad{0.0} \\
  w/o PHL &70.96 \downbad{6.24} &70.86 \downbad{6.60} &71.35 \downbad{8.63} &70.96 \downbad{6.24}\\
  w/o PHL, w/ HGNN &72.70 \downbad{4.50} &72.54 \downbad{4.92} &73.03 \downbad{6.95} &72.70 \downbad{4.50}\\
  w/o Inquiry Emb. & 72.63 \downbad{4.57} & 73.39 \downbad{4.07} & 74.22 \downbad{5.76} & 74.22 \downbad{2.98}\\

\bottomrule
\end{tabular}}
\label{tab:ablation_study}
\vspace{-0.1in}
\end{table}

We are also curious about the contribution of each component in \name contributes to the enhancements of \name. 
As shown in Table \ref{tab:ablation_study}, we successively removed the proposed prompt-attentive hypergraph learning (PHL) module, substituted the PHL module with HGNN~\cite{hgnn}, and removed the LLM-based argumentation. Note that this ablation study is conducted under hyperparameters setting different from those used for the results in Table \ref{tab:main}.

It can be observed from the experimental results in Table \ref{tab:ablation_study} that the original framework of our proposed \name delivers the best performance on verifying the factual knowledge stored in structured data. Firstly, removing the PHL module and directing the semantic embeddings directly to the projector results in a 6.24\% reduction in accuracy. Furthermore, to examine the role of hypergraph neural networks in enhancing LLMs' comprehension of structured knowledge, we replace our proposed PHL module with the classical HGNN~\cite{hgnn}, leading to a 4.50\% decrease in performance compared to \name, as shown in the third row of Table \ref{tab:ablation_study}. This performance degradation highlights the effectiveness of hypergraphs in representing structured knowledge. Specifically, We attribute this decline to the inability of HGNN to adequately leverage the information encoded in hyperedges for node updates during propagation. Additionally, we explore the impact of incorporating the inquiry embedding in the PHL module. As demonstrated in the last row of Table \ref{tab:ablation_study}, removing the inquiry embedding causes a substantial 5.76\% drop in precision and a more moderate 2.98\% decline in recall. This suggests that incorporating inquiry embeddings helps LLMs mitigate the bias toward over-generating positive responses, fostering more cautious reasoning by integrating the essential inquiry within prompts when processing structured data.

\vspace{-0.05in}
\subsection{Scalability (RQ5)}
\begin{figure}[t!]
    \centering
  \begin{minipage}[t]{0.49\linewidth} %
    \subfigure[Precision]{
        \includegraphics[width=\linewidth]{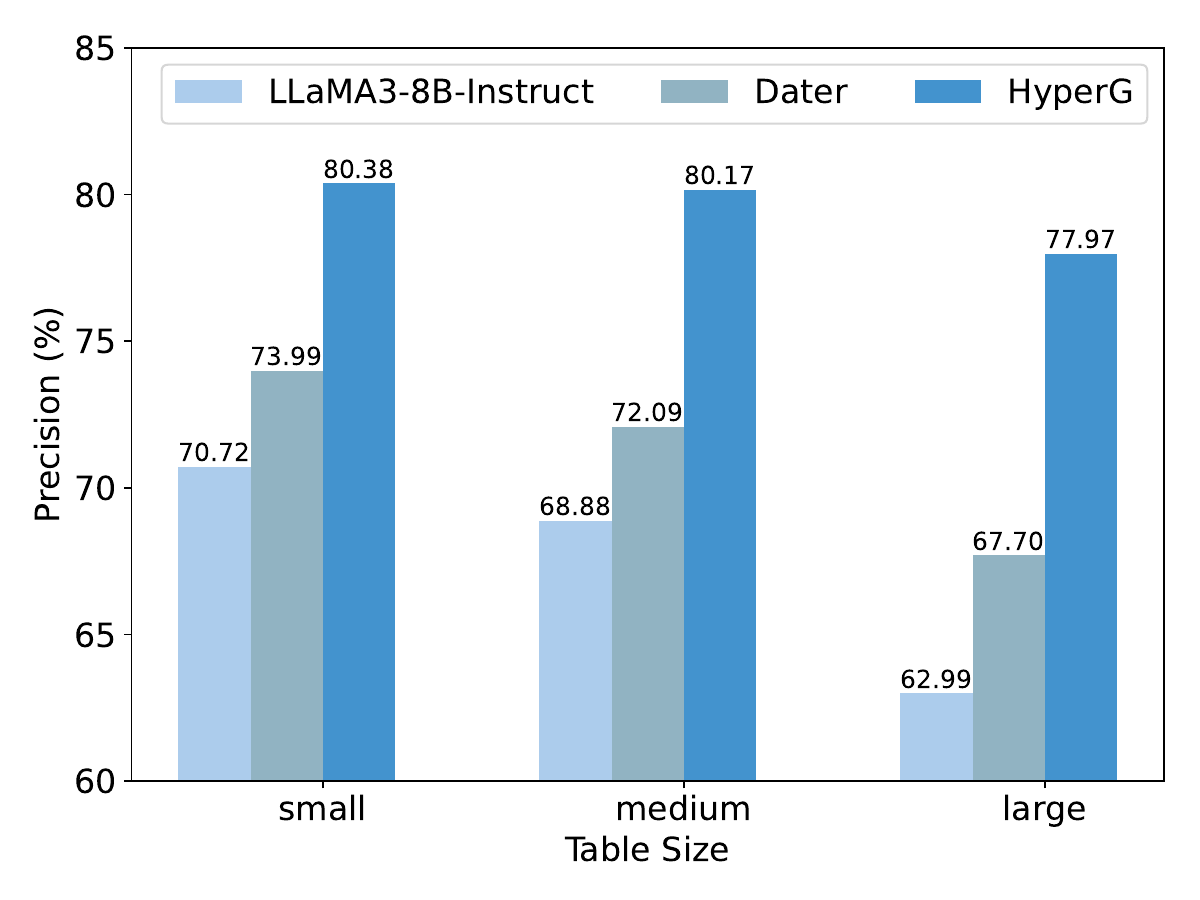}
    }  
  \end{minipage}
  \begin{minipage}[t]{0.49\linewidth} %
    \subfigure[Recall]{
        \includegraphics[width=\linewidth]{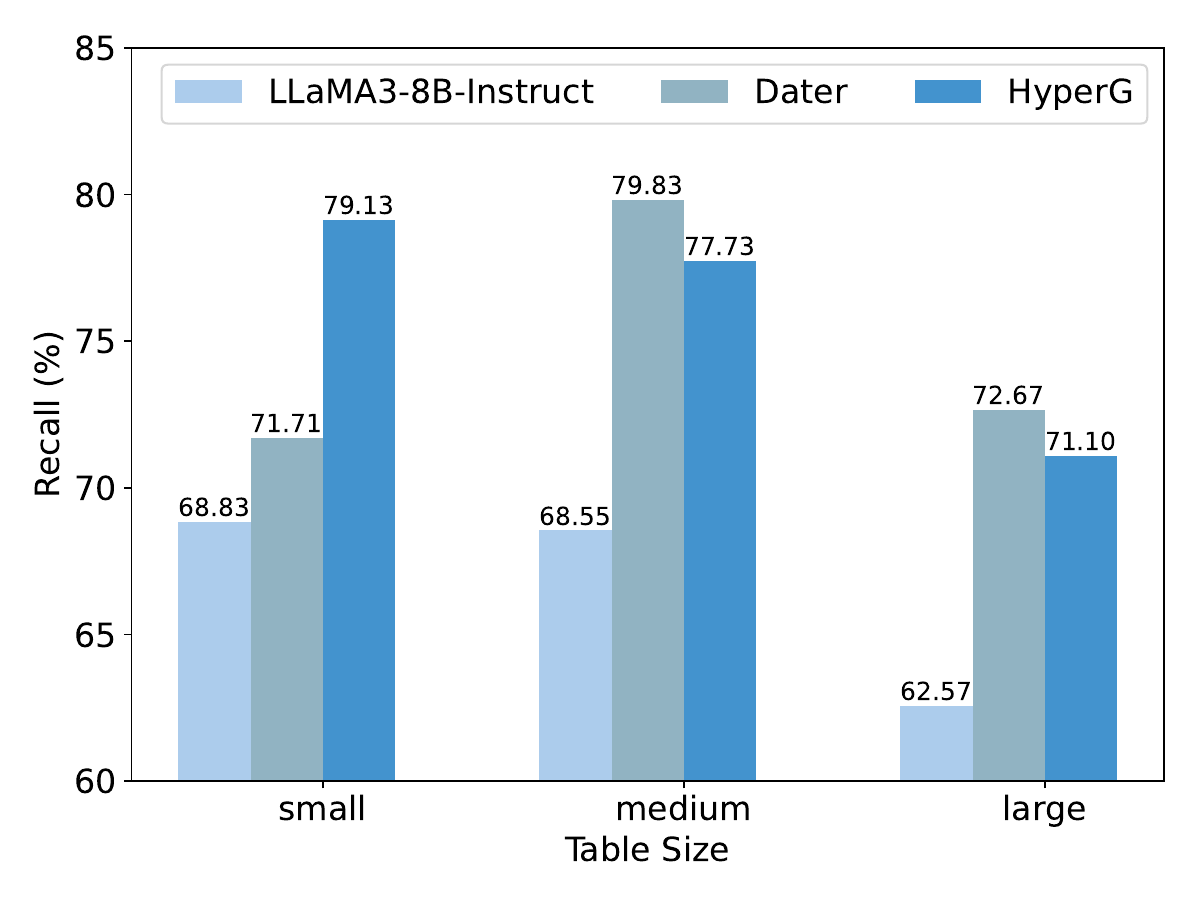}
    }  
  \end{minipage}
\vspace{-0.1in}
\caption{Performances of \name on tables of different sizes.}
\vspace{-0.2in}
\label{fig:scalability}
\end{figure}
Scalability is a critical concern, as large tables pose significant challenges for LLMs, which often struggle to interpret and integrate context from lengthy prompts~\cite{liu-etal-2024-lost,10.1145/3539618.3591708}. To evaluate the performances of \name on tables of varying sizes, we divided the testing tables in \cite{2019TabFactA} into three classes: \textbf{small} (\#rows $\leq$ 5 and \#columns$\leq$ 5), \textbf{medium} (6$\leq$\#rows$\leq$10 and 6$\leq$\#columns$\leq$ 10), and \textbf{large} (\#rows$\geq$ 10 and \#columns$\geq$ 10). We compare the performance of LLaMA3-8B-Instruct~\cite{touvron2024llama3}, Dater~\cite{dater}, and \name, with LLaMA3-8B-Instruct serving as the backbone model for all.

As illustrated in Figure \ref{fig:scalability}, with the table sizes increases there are generally declines in all of the models. \name demonstrates relatively stable performance in precision with a variance of 8.84, surpasses LLaMA3-8B-Instruct of 10.87. For small tables, \name surpasses Dater in precision and recall by 6.39\% and 7.42\%, respectively. However, Dater demonstrates superior recall performance compared to \name for medium and large tables. Upon careful examination and analysis of the true positive cases, we found that this is primarily due to the LLMs in Dater being inclined to generate positive answers. This is also benefits from its effective approach of decomposing queries and tables into sub- questions and sub-tables. The declines of \name in larger tables are primarily observed in recall. We attribute this to the limited number of hypergraph learning layers and aim to address this issue in the future through more sophisticated graph techniques.

\vspace{-0.05in}
\section{Related Works}
% structured knowledge + llms
% 1) serialized methods 
% 2) sql-based queries
Large Language Models (LLMs) excel in a broad spectrum of natural language tasks but face challenges when processing structured knowledge, such as tabular data~\cite{zhao-etal-2023-investigating}. Prior efforts to enhance LLMs' capabilities in handling structured knowledge can be broadly categorized into two main approaches: serialization-based methods~\cite{min2024exploring,hegselmann2023tabllm,jaitly2023towards} and operation-based methods~\cite{li2023sheetcopilot,ye2023decomposers,jiang-etal-2023-structgpt,wang2024chainoftable,lu2023chameleon}. 
Serialization-based methods convert structured data into a linear sequence of tokens, similar to how unstructured textual data is formatted for input into LLMs. TableLlama~\cite{zhang2024tablellama}, a pioneering approach to enhancing LLMs' performance on tabular data, is fine-tuned on the proposed TableInstruct dataset, which comprises serialized tables and task-specific instructions for several representative tabular tasks. %Similarly, GraphWiz~\cite{chen2024graphwiz} is trained on serialized graph data to improve performance on complex graphs, where the edges between nodes are explicitly enumerated. 
However, when dealing with highly complex tables or graphs, inquiry-relevant knowledge may be overlooked within the excessively long serialized token sequences~\cite{zhang2023ho,li2024snapkv}. 
The second category of methods resort to one or a series of operations such as SQL queries to help LLMs reason over structured data~\cite{li2023sheetcopilot,ye2023decomposers,jiang-etal-2023-structgpt,wang2024chainoftable,lu2023chameleon}. For example, Chain-of-Table~\cite{wang2024chainoftable} iteratively samples operations to select specific portions of the table that are tailored to the inquiry. Dater~\cite{ye2023decomposers} transforms the sub-questions generated by CodeX~\cite{chen2021evaluating} into SQL queries, enabling step-by-step multi-hop reasoning. Although these operation-based methods effectively locate the inquiry-relevant knowledge from structured data, they struggle when the target cell or neighboring cells contain missing or incomplete information.

%As messages propagate through the structures in Graph Neural Networks (GNNs), they naturally present themselves as a promising solution for enhancing the capabilities of Large Language Models. 
As messages propagate through the structures in Graph Neural Networks (GNNs), efforts have been made to integrate GNNs with LLMs to address structured knowledge more effectively~\cite{10.1145/3589334.3645627,ren2024survey,chai2023graphllm,tian2024graph,liu2024git}. For example, ~\citet{chai2023graphllm} uses a transformer module to encode the structured knowledge in graphs as the prefix of inputs to the LLMs.
Additionally, graphs serve as powerful tools for representing tabular data~\cite{chen2024hytrel,jin2024hgt}. HGT~\cite{jin2024hgt} explicitly models tables as graphs by connecting various components within the tables to enhance LLM capabilities. %Although effective, constructing this type of heterogeneous graph requires manual annotation of various table components, and the relevance of headers to the cells tends to diminish as the table length increases, thereby harming the hierarchical dependencies. 
Furthermore, HYTREL~\cite{chen2024hytrel} is particularly relevant to our \name as it also empolys hypergraphs to represent tabular data, but it overlooks incorporating the semantics of task within prompts during message propagation. Existing works, while effective, primarily focus on utilizing LLMs rather than improving their inherent capabilities with model-agnostic modules. To the best of our knowledge, we are the first to leverage hypergraphs to enhance the capabilities of LLMs in handling structured knowledge.

% \vspace{-0.18in}
\section{Conclusion}
In this paper, we present a novel hypergraph-based generation framework, \name, designed to enhance the understanding and reasoning capabilities of Large Language Models (LLMs) when dealing with knowledge structurally stored. The primary objective of \name is to tackle the challenges arising from complex structural relationships and data sparsity, such as incomplete cell information, within structured data. By employing a novel prompt-attentive hypergraph learning (PHL) module, \name effectively propagates information across high-order group dependencies, capturing intricate connections within the data. Comprehensive experiments across three distinct tabular tasks consistently demonstrate the impact of \name on enhancing the performance of LLMs with different parameter scales. We envision \name as a solution for enhancing LLMs in a broader range of applications which requires nuanced understanding of structured information. Furthermore, \name facilitates the broader adoption of LLMs in real-world applications, where knowledge is often stored in structured formats, thereby enabling LLMs to better handle and reason over such data.

% \section{Appendices}

% \begin{acks}
% \end{acks}

% \bibliographystyle{ACM-Reference-Format}
% \bibliography{ref}
\clearpage









\end{document}